\documentclass[10pt]{article}
\usepackage{graphicx}
\usepackage{amsmath}

\def\Title#1{\begin{center} {\Large #1 } \end{center}}
\def\Author#1{\begin{center}{ \sc #1} \end{center}}
\def\Address#1{\begin{center}{ \it #1} \end{center}}

\newcommand\pubblock{\rightline{\begin{tabular}{l} Proceedings of the Second Annual LHCP\\ \pubnumber\\
         \pubdate  \end{tabular}}}

\newenvironment{Abstract}{\begin{quotation} \begin{center} 
             \large ABSTRACT \end{center}\bigskip 
      \begin{center}\begin{large}}{\end{large}\end{center} \end{quotation}}

\newenvironment{Presented}{\begin{quotation} \begin{center} 
             PRESENTED AT\end{center}\bigskip 
      \begin{center}\begin{large}}{\end{large}\end{center} \end{quotation}}

\def\beq{\begin{equation}}
\def\eeq#1{\label{#1}\end{equation}}
\def\eeqn{\end{equation}}

\def\beqa{\begin{eqnarray}}
\def\eeqa#1{\label{#1}\end{eqnarray}}
\def\eeqan{\end{eqnarray}}

\let\bar=\overbar

\def\Dslash{\not{\hbox{\kern-4pt $D$}}}
\def\dslash{\not{\hbox{\kern-2pt $\del$}}}

\def\msb{{\bar{\ssstyle M \kern -1pt S}}}

\usepackage{color}

\newcommand\pubnumber{ CMS CR-2014/207 }
\newcommand{\etmiss}{\mathrm{E}_{\mathrm{T}}^{\mathrm{miss}}}

\newcommand\pubdate{\today}

\def\affiliation{
On behalf of the CMS Experiment, \\
Department of Physics \\
University of California, Los Angeles, CA 90095, U.S.A }

\begin{document}

\large
\begin{titlepage}
\pubblock

\vfill
\Title{Search for electroweak SUSY production at CMS}
\vfill

\Author{ Pieter Everaerts }
\Address{\affiliation}
\vfill
\begin{Abstract}
Using the data collected during Run I of LHC operation the CMS Collaboration performed multiple analyses searching for the direct electroweak production of supersymmetric particles in proton-proton collisions.   Different decay modes of the gauginos and sleptons were considered, through intermediate vector bosons or Higgs bosons, or directly to leptons.  A set of complementary searches were designed to target these different decays.  None of these searches shows any indication for physics beyond the standard model.

\end{Abstract}
\vfill

\begin{Presented}
The Second Annual Conference\\
 on Large Hadron Collider Physics \\
Columbia University, New York, U.S.A \\ 
June 2-7, 2014
\end{Presented}
\vfill
\end{titlepage}
\def\thefootnote{\fnsymbol{footnote}}
\setcounter{footnote}{0}
%

\normalsize 


\section{Introduction}

Most searches for supersymmetry (SUSY) at the LHC focused on the strong production of supersymmetric particles because of its large production cross section.  Thanks to these studies, gluinos and inclusive squarks have been probed up to masses above 1 TeV.  Electroweak production of SUSY has in general smaller cross sections, but the production cross sections for winos, binos or higgsinos, collectively referred to as electroweakinos, of around 300--400 GeV are similar to those of squarks and gluinos with masses around 1--1.5 TeV.  This note focuses on the searches for direct electroweak production in pp collisions at the LHC performed by the CMS~\cite{cms} collaboration~\cite{ewkino,fourb}.  The pair production of gauginos and sleptons leads to very clean experimental signatures due to the fact that the electroweakinos decay to leptons, vector bosons or Higgs bosons.  The only hadronic activity in the event is due to bosonic decay products and initial-state radiation.  Therefore top-related background can be heavily suppressed by applying a veto on the amount of hadronic activity and/or the presence of b-tagged jets in the event.  The other major background to these searches is dibosons.  Those are suppressed by using the missing transverse energy ($\etmiss$) and other kinematic variables (e.g. transverse mass).  For the signal models without intermediate vector bosons, events with Z candidates are also vetoed. 

This note mainly focuses on chargino-neutralino pair production, but also gives the highlights of the other CMS searches for electroweak production of SUSY.  If the sleptons and sneutrinos are light, then the chargino and neutralino decay through intermediate sleptons and sneutrinos (Sec.~\ref{charginoneutralinoslepton}).  If they are heavy, then the decay happens through intermediate vector bosons (Sec.~\ref{charginoneutralinovector boson}) and Higgs bosons (Sec.~\ref{charginoneutralinoHiggs}). Section~\ref{neutralinopair} discusses the searches for neutralino pair production.  Finally the results for chargino pair production and slepton pair production are shown in Sec.~\ref{charginoslepton}.   

\section{Chargino-neutralino pair production with intermediate sleptons and sneutrinos} \label{charginoneutralinoslepton}

If the sleptons and sneutrinos are light, then neutralinos and charginos decay to the lightest supersymmetric particle (LSP) through intermediate sleptons and sneutrinos, giving rise to a final state with three leptons and two lighest supersymmetric particles (LSP).  This final state is targeted by the trilepton analysis, where Z candidates are explicitly vetoed.  In case of compressed mass spectra, however, those leptons can be very soft and escape detection.  A same-sign dilepton analysis is used to complement the trilepton analysis in these areas of SUSY parameter space.


\subsection{Trilepton analysis} \label{trilepton}

The trilepton analysis~\cite{ewkino} targets final states with exactly three leptons. 
The most sensitive search regions are those with an opposite-sign same-flavor lepton pair.  The main backgrounds to this analysis are WZ and top pair production.  To suppress the WZ production and other backgrounds with real Zs present, the sample is split between a sample with and a sample without a Z candidate.  The first sample is used for the chargino-neutralino decays with intermediate vector bosons (Sec.~\ref{charginoneutralinovector boson}), the second one is the most sensitive for the decays with intermediate sleptons and sneutrinos.  The WZ background is further suppressed using the transverse mass variable, calculated with the $\etmiss$ and the transverse momentum of the third lepton.  Top quark backgrounds are reduced by applying a veto on the presence of b-tagged jets in the event.  Also $\etmiss$ binning is to create search regions with lower standard model (SM) backgrounds.    


The backgrounds due to non-prompt or misidentified leptons are estimated using data-driven techniques that measure the probability of a non-prompt lepton to be isolated in a control sample.  Simulation is used to estimate the diboson backgrounds and other rare processes.  For the WZ background, the simulation of the hadronic recoil and the $\etmiss$ is corrected by making detailed comparisons between data and simulation in Z+jets events.  Using these background predictions no evidence for new physics is found in the trilepton analysis.

\subsection{Same-Sign dilepton analysis}

Compressed mass spectra can lead to very soft leptons, which can then escape detection.  These final states can be targeted with dilepton final states.  A search using opposite-sign dileptons would suffer from large standard model backgrounds, but those are strongly reduced if we ask for same-sign dileptons instead~\cite{ewkino}. If one of the leptons from the neutralino is not detected or identified, then there is a 50\% chance to end up with same-sign dileptons.  Compared to the trilepton analysis, the contribution of the backgrounds due to non-prompt and misidentified leptons is strongly enhanced.   To counteract this effect, the requirements on the lepton transverse momenta, the $\etmiss$, or the hadronic activity can be tightened.     All these different strategies are used.  The same-sign analysis asks for larger lepton transverse momenta than the trilepton analysis.  One search region applies very tight requirements on the $\etmiss$ ($>$200 GeV) and also vetoes events with b-tagged jets to reduce the top backgrounds.  The other search region has a more moderate  $\etmiss$ requirement (120$<\etmiss<$200 GeV), but then has a tighter requirement for the hadronic activity, asking for no jets at all to be present in the event.     

The background prediction methods are very similar to those for the trilepton analysis.  The non-prompt and misidentified leptons are predicted with a data-driven method using the control region with one non-isolated lepton, and the diboson backgrounds are estimated with simulation that has been cross-checked in diboson-enriched control regions.  Good agreement has been observed between the observed and predicted yields.

\subsection{Interpretation}

The trilepton and same-sign dilepton analysis can be interpreted in terms of chargino-neutralino pair production with both gauginos decaying to a LSP through intermediate sleptons and sneutrinos~\cite{ewkino}.  The chargino and neutralino are assumed to have the same mass.  Since the same-sign dilepton analysis provides results with an explicit third lepton veto, the results of the same-sign dilepton analysis and trilepton can be combined. 

\begin{figure}[htb]
\centering
\includegraphics[width=0.4\textwidth]{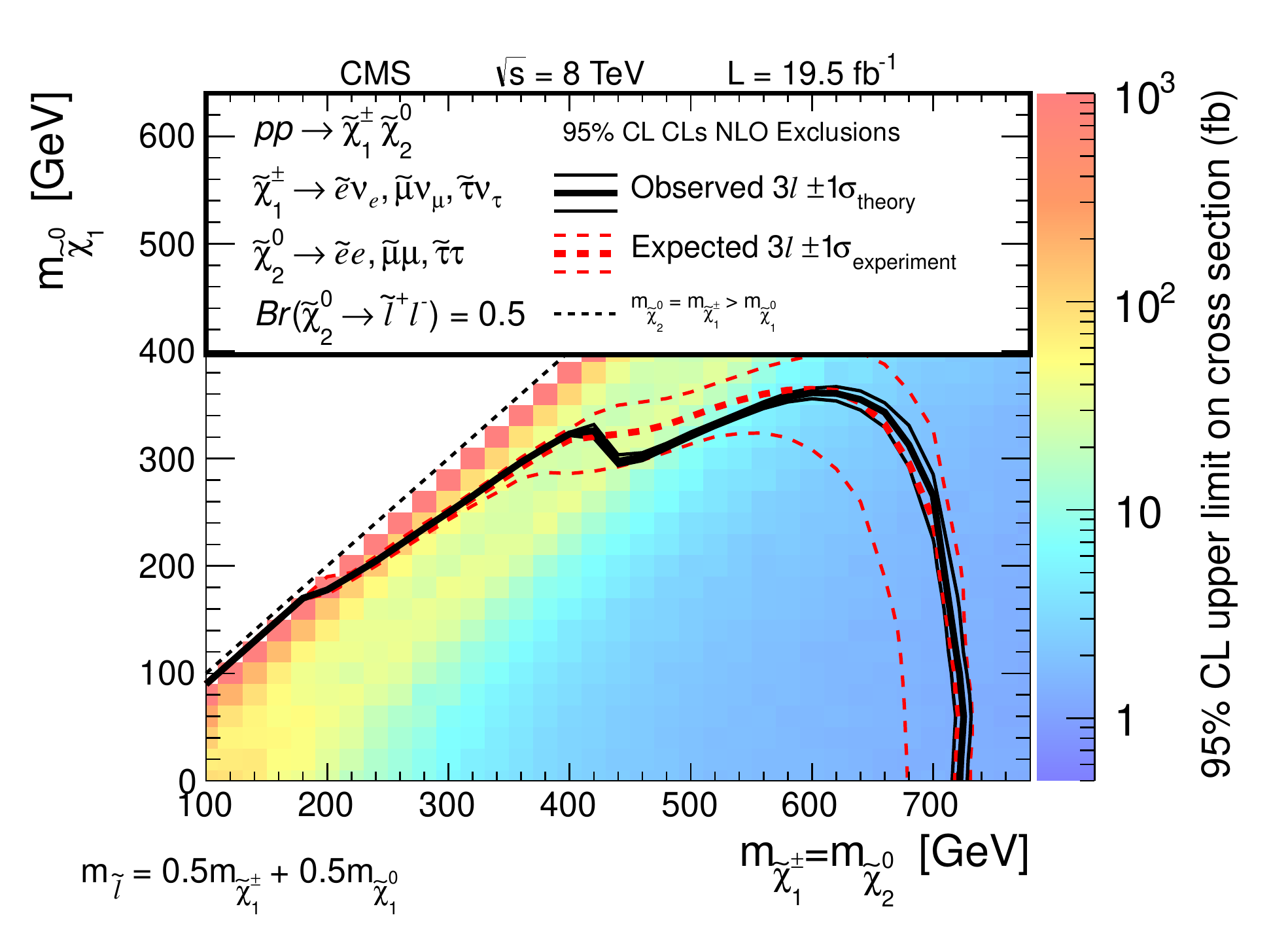}
\includegraphics[width=0.4\textwidth]{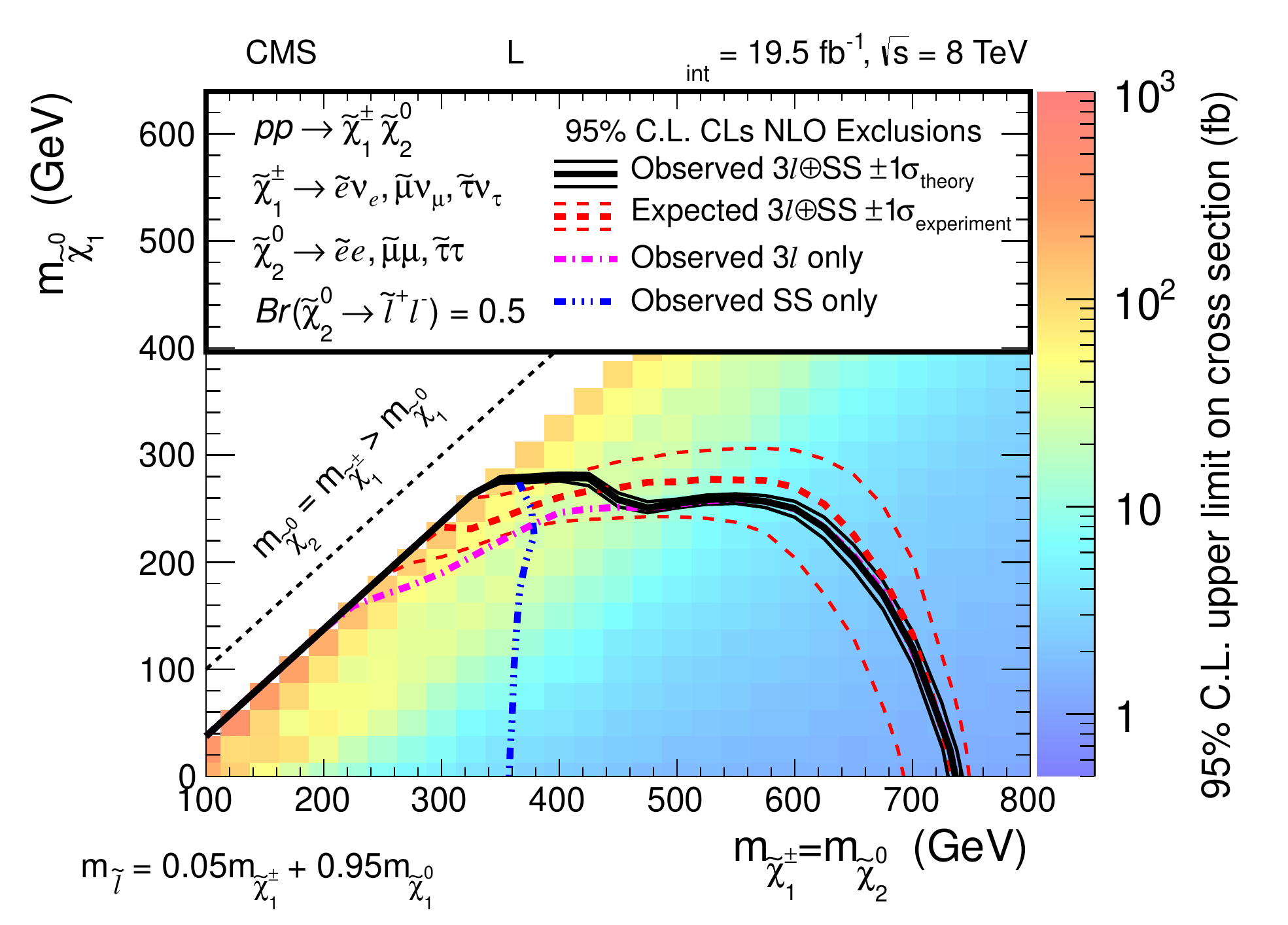}
\caption{Interpretation of the results of the three-lepton and same-sign dilepton search in the flavor-democratic signal
model with slepton mass in the middle between the chargino and LSP mass (left) and very close the the LSP mass (right,$m_{\tilde{l}}=0.95m_{\tilde{\chi}_1^0}+0.05m_{\tilde{\chi}^\pm}$).}  
\label{chargneut_slep}
\end{figure}

Figure~\ref{chargneut_slep} shows the interpretation results for the chargino-neutralino pair production.  The 95\% CL upper limit on the cross section times branching fraction is shown as a function of the masses of the chargino and the LSP.  The two plots show different possibilities for the LSP mass.  The left plot of Fig.~\ref{chargneut_slep} assumes that the slepton mass is exactly in the middle between the chargino and the LSP mass, while the right plot shows a more compressed mass spectrum where the slepton mass is very close to the LSP mass. ($m_{\tilde{l}}=0.95m_{\tilde{\chi}_1^0}+0.05m_{\tilde{\chi}^\pm}$).  The upper limit on the cross section times branching fraction generally becomes more stringent with the increasing mass difference between the chargino or heavy neutralino and the LSP. A reduced sensitivity is observed in the region where this mass difference is consistent with the invariant mass of the Z boson, and is caused by a higher rate for the WZ background.   Chargino and neutralino masses are probed up to 720 GeV.  The complementarity of the same-sign dilepton analysis can be seen in the right plot of Fig.~\ref{chargneut_slep} where these results increase the excluded region of parameter space towards the smaller mass splittings between the chargino and the LSP.  

\section{Chargino-neutralino pair production with intermediate vector bosons} \label{charginoneutralinovector boson}

If all the sleptons and sneutrinos are heavy, the chargino and neutralino decay predominantly through vector bosons and Higgs bosons.  This section focuses on decays with intermediate vector bosons while Sec.~\ref{charginoneutralinoHiggs} discusses decays with Higgs bosons.  The experimental signature for the decay through vector bosons is WZ+$\etmiss$. The W and Z bosons then decay according to their standard model branching fractions into quarks and leptons.  CMS targets this signature with two analyses.  In both cases the Z is assumed to decay leptonically but the W can decay leptonically or hadronically.  The fully leptonic decay leads to a final state with three leptons.  The trilepton analysis (Sec.~\ref{trilepton}) is used to target these final states, using search regions both with and without a Z candidate.  If the W decays hadronically, the final state has a dilepton pair consistent with the Z boson and a dijet pair consistent with a W boson.  

\subsection{Z + dijet analysis} \label{Zdijet}

The Z boson is required to decay leptonically and to be fully reconstructed, while the hadronic decay of the W boson is identified by its dijet signature~\cite{ewkino}.  This is done by requiring the invariant mass of the dijet to be consistent with the mass of a W boson.  Events with b-tagged jets are vetoed to reduce the top backgrounds.   Using the $\etmiss$ shape, the electroweakino signal can be distinguished from the dominant Z + jets background.  The $\etmiss$ spectrum is split into different bins because the $\etmiss$ in the SUSY events depends on the exact chargino/neutralino and LSP masses.  A template fit is performed with a Z+jets $\etmiss$ template shape derived from a $\gamma$ + jets control sample.  Flavor-symmetric backgrounds with real $\etmiss$ such as top pair production and Z$\rightarrow\tau\tau$ are estimated from an e$\mu$ control region. The other backgrounds are estimated using simulated samples.  The observed and predicted $\etmiss$ spectra agree over several orders of magnitude. 

\subsection{Interpretation}

\begin{figure}[htb]
\centering
\includegraphics[width=0.4\textwidth]{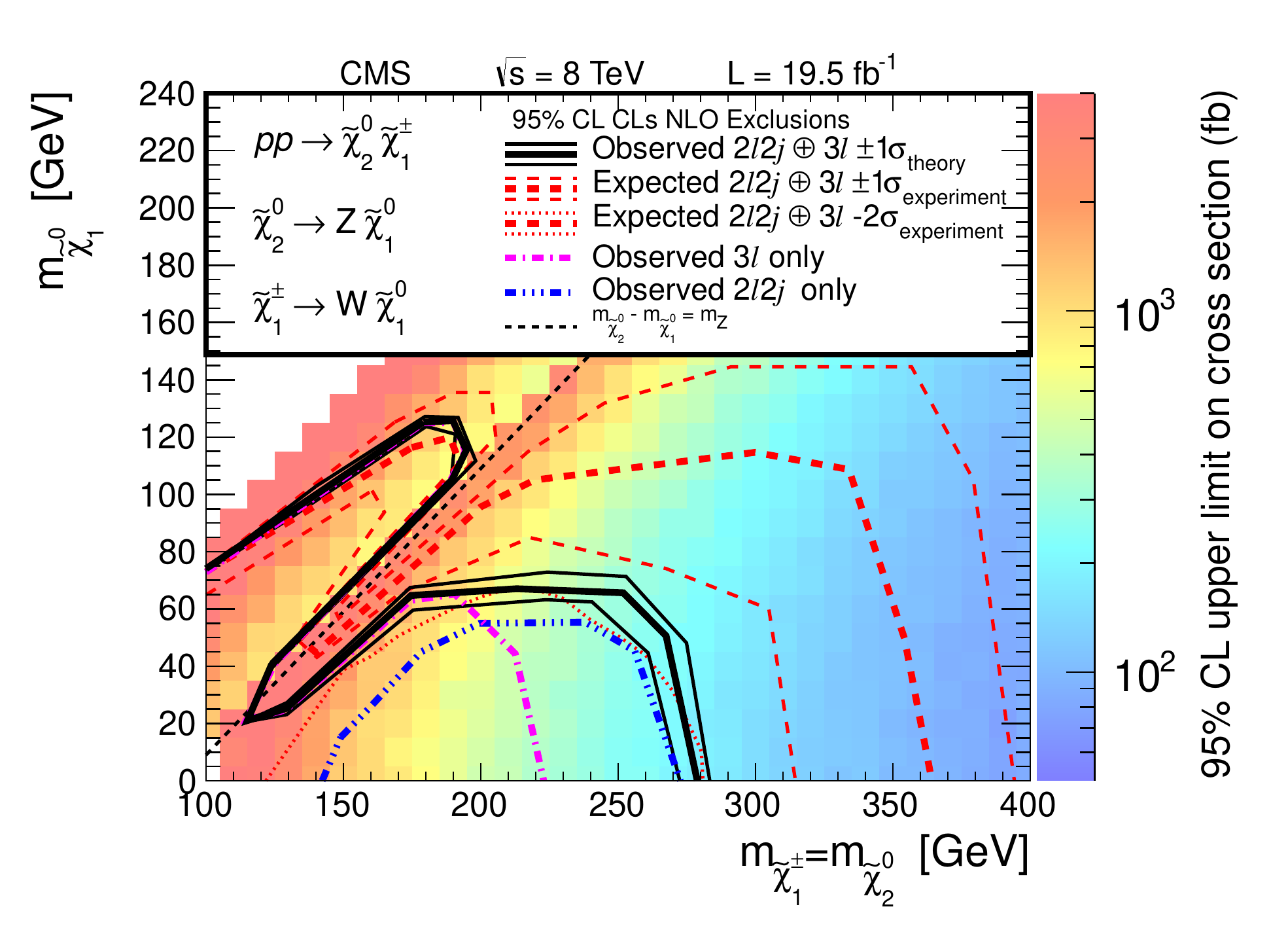}
\caption{Interpretation of the results of the three-lepton and Z+dijet search for chargino-neutralino production with intermediate WZ in the decay.} 
\label{chargneut_vector}
\end{figure}

Figure~\ref{chargneut_vector} shows the results of the trilepton and the Z+dijet analysis for chargino-neutralino pair production where the chargino and neutralino decay through W and Z bosons~\cite{ewkino}.   The Z+dijet analysis is the most sensitive for large chargino and small LSP masses, while the trilepton analysis gives the strongest exclusion limits for the more compressed spectra.  The trilepton search regions with a Z candidate have the largest sensitivity if the mass difference is larger than the Z mass, while the search regions without a Z candidate are the most important for the even smaller mass splittings.  A significant degradation in sensitivity is present in the region of parameter space where the mass difference is almost equal to the Z mass, because the chargino and neutralino decay products are then produced with low momentum in the rest frame of their parent particles. Chargino masses up to 270 GeV are probed by the combination of the two analyses.

\section{Chargino-neutralino pair production with intermediate vector bosons and Higgs bosons} \label{charginoneutralinoHiggs}

In this section we assume that the branching fraction of the neutralino decay to a Higgs boson and a LSP is 100\%.  Then chargino-neutralino pair production has a WH+$\etmiss$ signature.  For the signal model we assume a Higgs boson with a mass of 126 GeV and the SM branching fractions.  The W  is required to decay leptonically.  The different analyses target the dominant Higgs boson decays.  The largest branching fraction of the Higgs boson decay is the b$\bar{\text{b}}$ final state and this one is targeted by a single-lepton analysis.  The requirements of the same-sign analysis are tuned to be sensitive to the H$\rightarrow$WW decay.  Finally the generic multilepton analysis is also re-interpreted to give sensitivity to the possible H$\rightarrow$WW, H$\rightarrow$ZZ and H$\rightarrow\tau\tau$ decays.

\subsection{Single-lepton analysis}

The single-lepton analysis~\cite{ewkino} assumes that the Higgs boson decays to a pair of b-jets and the W decays to a lepton and a neutrino.  The two LSPs in the event give extra $\etmiss$ and this is used by applying tight requirements on $\etmiss$-related kinematic variables, such as the transverse mass, and also using a $\etmiss$ binning.  There is also a requirement to have exactly two jets in the event, and both of them have to be b-tagged.  This helps to reduce top-related background, which would have additional jet activity in the event.  The main backgrounds, like top pair production and W+jets production, are estimated using simulation with data-driven corrections that are derived in a set of control regions.  Finally the Higgs boson mass is exploited by looking for a resonance in the mass spectrum of the b$\bar{\text{b}}$ dijet pair.  

\subsection{Same Sign dilepton analysis}

The same-sign dilepton analysis~\cite{ewkino} targets the H$\rightarrow$WW decay.  It asks for exactly two high $p_T$ leptons and two or three jets.  Not allowing more than three jets helps to limit rare backgrounds like top pair production in association with a vector boson.  Events with b-jets are vetoed to suppress top pair production.  Similar to the single-lepton analysis, $\etmiss$  and $\etmiss$-related kinematic variables are used to suppress the SM backgrounds further.   Non-prompt and misidentified leptons are estimated in a data-driven way by looking at a control region where one of the leptons is non-isolated.  The other backgrounds are estimated using simulated samples.  After applying these requirements, the visible mass of the Higgs boson candidate is investigated.  This is done by calculating the mass of the lepton plus 2-jet system.  The visible mass is below the Higgs boson mass for signal, while most SM backgrounds yield a large value for this mass variable.  

\subsection{Multilepton analysis}

The results of the CMS inclusive multilepton analysis~\cite{multilepton} are also reinterpreted in this model.  This analysis looks at events with three or four leptons of which at most one can be a hadronically decaying tau lepton.  Then a detailed binning is performed in the number and type of leptons, the number of b-jets, the hadronic activity and the $\etmiss$.  The predictions for Z+jets and Z$\gamma$ are made using data-driven methods, while the other backgrounds are estimated with simulation.  The simulation is corrected by looking at background-enriched control regions.  For the electroweakino models, only the regions without any b-jets and with low hadronic activity are relevant.  In those regions of phase space no excess is observed.

\subsection{Interpretations}
Chargino masses are probed up to 200 GeV if the chargino and neutralino decay 100\% through the combination of a W and a Higgs boson~\cite{ewkino}.  The single-lepton analysis is the most powerful analysis for the large chargino masses, but at small chargino masses the combination with the same-sign dilepton and multilepton analysis yields a small improvement in the calculated upper limit, which increases the mass reach by 50 GeV.



\section{Neutralino pair production} \label{neutralinopair}
As discussed in Sec.~\ref{charginoneutralinovector boson}, the next-to-lightest neutralino can decay to the LSP by emitting a Z or a Higgs boson.  In the chargino-neutralino pair production, this results in both WZ+$\etmiss$ and WH+$\etmiss$  signatures.  In the case of neutralino pair production the decay can happen through two Z bosons, two Higgs bosons, or one Higgs and one Z boson.  The exact combination of these signatures depends on the branching fractions of the neutralino decay.  In this note we consider the two extreme cases: all neutralinos decay through Higgs bosons, or all neutralinos decay through Z bosons. The neutralino pair production cross section can be enhanced in a gauge-mediated symmetry breaking supersymmetric model.  In this case the gravitino is the LSP and the next-to-lightest supersymmetric particle is a neutralino. 

If the neutralinos decay through Z bosons, then those Z bosons can decay leptonically or hadronically.  To suppress the SM background, one of the two Zs is required to decay leptonically and be fully reconstructed.  The other Z boson can decay in either way.  In case of the hadronic decay of the Z boson, the signature will be Z+dijet with the dijet pair having a mass close to the Z boson.  The Z+dijet analysis of Sec.~\ref{Zdijet} can be used to look for this final state.  For the leptonic decay of the Z boson, a search in the four lepton final state is performed.  To target neutralino pair production with decays through two Higgs bosons, an analysis is developed that looks in the four b-jet final state.  
\subsection{Four lepton analysis}

The four-lepton analysis~\cite{ewkino} uses bins in the number of opposite-sign same-flavor pairs, the dilepton mass of these pairs and also the $\etmiss$ of the event.  The methods used are very similar to the methods in the trilepton analysis (Sec.~\ref{trilepton}).  The non-prompt and misidentified leptons are estimated using a data-driven method based on the lepton isolation, while the diboson backgrounds are predicted from simulation, using corrections derived from data control regions.  

\subsection{Four b-jet analysis}

The four b-jet analysis~\cite{fourb} targets neutralino pair production with decays to two Higgs bosons.  Higgs bosons predominantly decay to b$\bar{\text{b}}$, so the final state will have four b-jets and two LSPs.  Exactly four or exactly five jets are required to reduce rare SM backgrounds with high jet multiplicity.  A lepton veto and an additional charged track veto are applied to remove backgrounds with leptons.  SM backgrounds are further diminished by asking for a large $\etmiss$ significance and applying topological cuts on the angles between the jets and the $\etmiss$.  It is also required that at least three jets are b-tagged.  The jets are paired into two Higgs boson candidates by looking at the smallest difference between the masses of the two resulting jet pairs.  The average jet pair mass is required to be in the Higgs boson mass window (100-140 GeV).  The background is estimated by extrapolating the main backgrounds from a set of control regions.  These control regions for this estimate are made by requiring only two b-tagged jets and by looking at regions where the jet pair mass constraints are not fulfilled.

\section{Slepton and chargino pair production} \label{charginoslepton}

Slepton and chargino pair production~\cite{ewkino} both lead to signatures with two opposite-sign dileptons and LSPs giving large $\etmiss$.  In the chargino case, there are extra neutrinos present in the event but the $\etmiss$ from those is expected to be considerable smaller than from the LSPs and does not influence the signature much.  In the slepton case the two leptons have the same flavor because the sleptons decay to a specific lepton flavor, while the two charginos decay independently to both lepton flavors and are thus just as likely to have an opposite-flavor as a same-flavor pair. The events are split based on same-flavor or opposite-flavor to exploit this difference.  To remove the large Z-related background, a dilepton mass constraint is added to the same-flavor dilepton pair.  The analysis applies a b-jet veto to reduce top pair background and a moderate $\etmiss$ cut is applied to suppress SM backgrounds.   The MC$_{\text{T,Perp}}$-variable (kinematic variable with endpoint for W-related backgrounds) is fitted with templates for the SM backgrounds.  Those templates are obtained from data control regions if possible or taken directly from simulation. Left-handed sleptons are probed up to 260 GeV and charginos up to 540 GeV.

\section{Summary}
\begin{figure}[htb]
\centering
\includegraphics[width=0.4\textwidth]{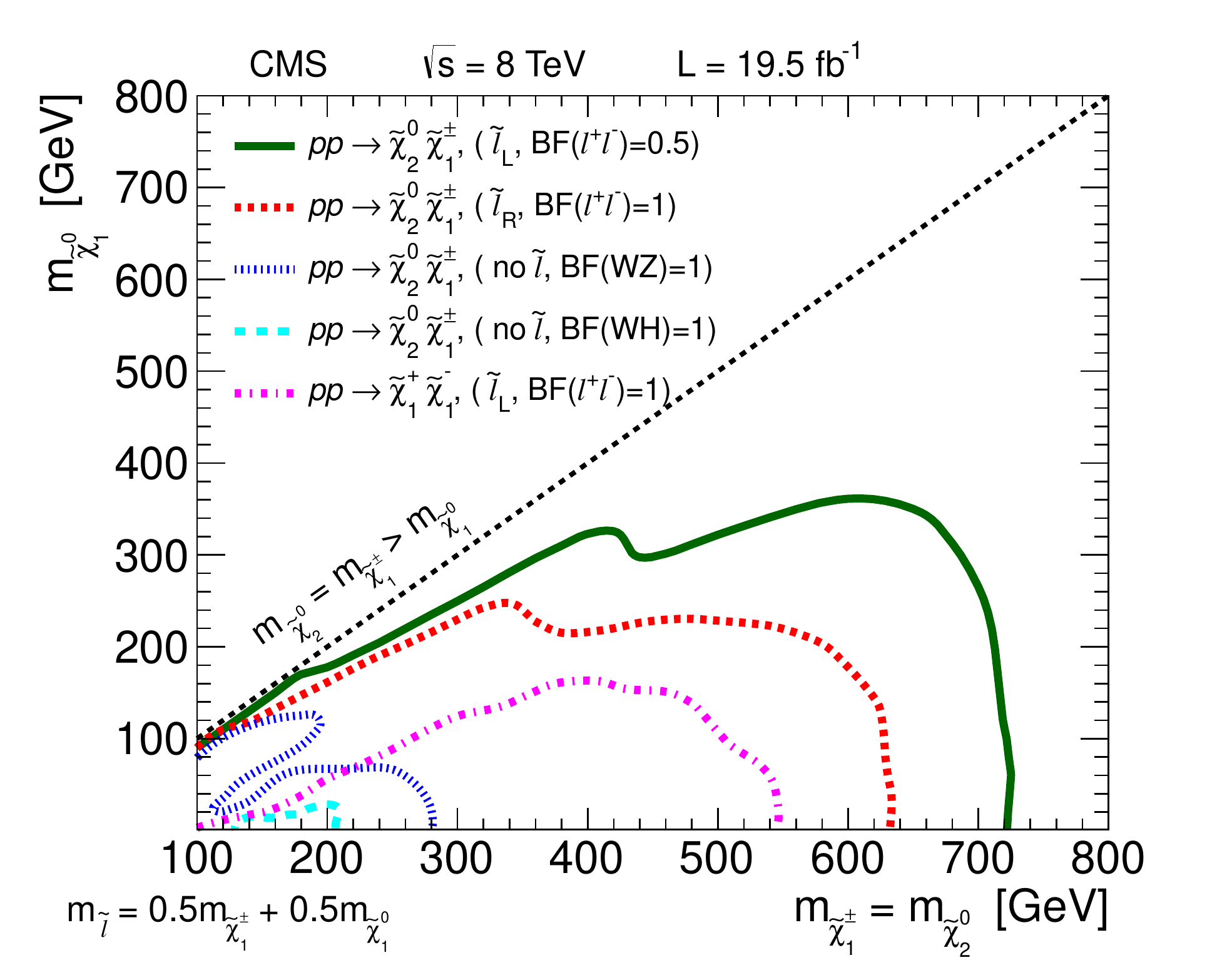}
\includegraphics[width=0.4\textwidth]{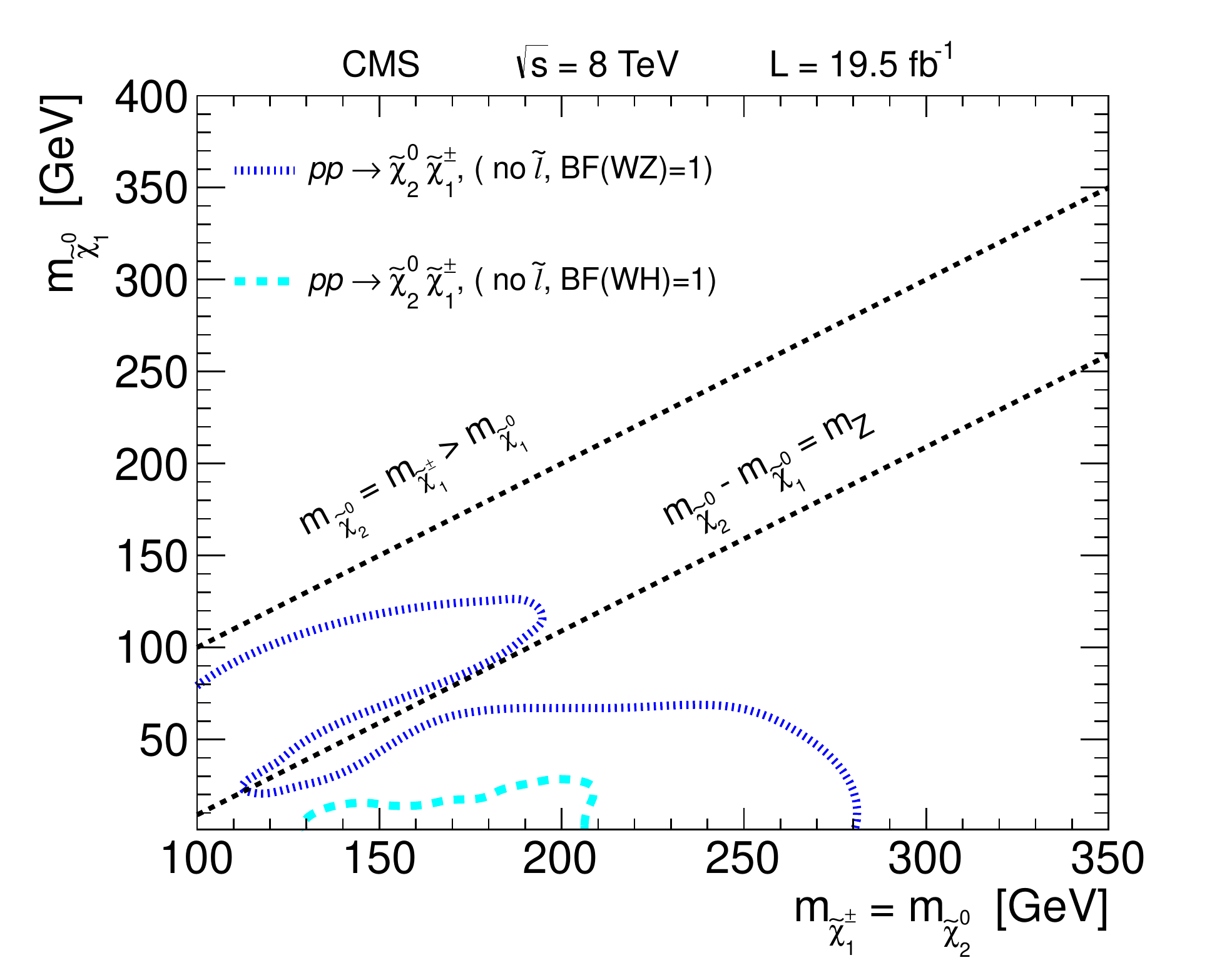}
\caption{  (left) Contours bounding the mass regions excluded at 95\% CL for chargino-neutralino production with decays to left-handed sleptons, right-handed sleptons, or direct decays to Higgs and vector bosons, and for chargino-pair production, based on NLO+NLL signal cross sections. (right) Expanded view for chargino-neutralino production with decays to Higgs and vector bosons.
}
\label{fig:summary}
\end{figure}

CMS performed a wide variety of searches for electroweak production of SUSY~\cite{ewkino,fourb}.  No evidence for new physics is found, but stringent limits are put on the production of such particles.  Depending on the exact decay mode the production of charginos and neutralinos is probed up to 200-700 GeV.  Figure~\ref{fig:summary} shows the summary of the exclusion results. In the left plot all the different models are shown, while the right one zooms in on the bosonic decays of chargino-neutralino production.  In general, the slepton-mediated decays are more powerful than the boson-mediated ones and the Higgs boson decays are the hardest to probe.  For example the slepton-mediated decay of chargino-neutralino pair production is probed up to 720 GeV, but this limit goes down to 280 GeV for the WZ-mediated decay and 210 GeV for the WH signature.

\end{document}